
%
%
%
\def\unredoffs{} \def\redoffs{\voffset=-.31truein\hoffset=-.59truein}
\def\speclscape{\special{ps: landscape}}
%
%
%
%
\newbox\leftpage \newdimen\fullhsize \newdimen\hstitle \newdimen\hsbody
\tolerance=1000\hfuzz=2pt
\catcode`\@=11 
\def\bigans{b }
\message{ big or little (b/l)? }\read-1 to\answ
\ifx\answ\bigans\message{(This will come out unreduced.}
\magnification=1200\unredoffs\baselineskip=16pt plus 2pt minus 1pt
\hsbody=\hsize \hstitle=\hsize 
\else\message{(This will be reduced.} \let\l@r=L
\magnification=1000\baselineskip=16pt plus 2pt minus 1pt \vsize=7truein
\redoffs \hstitle=8truein\hsbody=4.75truein\fullhsize=10truein\hsize=\hsbody
\output={\ifnum\pageno=0 
  \shipout\vbox{\speclscape{\hsize\fullhsize\makeheadline}
    \hbox to \fullhsize{\hfill\pagebody\hfill}}\advancepageno
  \else
  \almostshipout{\leftline{\vbox{\pagebody\makefootline}}}\advancepageno
  \fi}
\def\almostshipout#1{\if L\l@r \count1=1 \message{[\the\count0.\the\count1]}
      \global\setbox\leftpage=#1 \global\let\l@r=R
 \else \count1=2
  \shipout\vbox{\speclscape{\hsize\fullhsize\makeheadline}
      \hbox to\fullhsize{\box\leftpage\hfil#1}}  \global\let\l@r=L\fi}
\fi
%
\newcount\yearltd\yearltd=\year\advance\yearltd by -1900

%
%

\def\draftmode{\message{ DRAFTMODE }\def\draftdate{{\rm preliminary draft:
\number\month/\number\day/\number\yearltd\ \ \hourmin}}%
\headline={\hfil\draftdate}\writelabels\baselineskip=20pt plus 2pt minus 2pt
 {\count255=\time\divide\count255 by 60 \xdef\hourmin{\number\count255}
  \multiply\count255 by-60\advance\count255 by\time
  \xdef\hourmin{\hourmin:\ifnum\count255<10 0\fi\the\count255}}}
\def\nolabels{\def\wrlabeL##1{}\def\eqlabeL##1{}\def\reflabeL##1{}}
\def\writelabels{\def\wrlabeL##1{\leavevmode\vadjust{\rlap{\smash%
{\line{{\escapechar=` \hfill\rlap{\sevenrm\hskip.03in\string##1}}}}}}}%
\def\eqlabeL##1{{\escapechar-1\rlap{\sevenrm\hskip.05in\string##1}}}%
\def\reflabeL##1{\noexpand\llap{\noexpand\sevenrm\string\string\string##1}}}
\nolabels
%
\global\newcount\secno \global\secno=0
\global\newcount\meqno \global\meqno=1
\def\newsec#1{\global\advance\secno by1\message{(\the\secno. #1)}
\global\subsecno=0\eqnres@t\noindent{\bf\the\secno. #1}
\writetoca{{\secsym} {#1}}\par\nobreak\medskip\nobreak}
\def\eqnres@t{\xdef\secsym{\the\secno.}\global\meqno=1\bigbreak\bigskip}
\def\sequentialequations{\def\eqnres@t{\bigbreak}}\xdef\secsym{}
\global\newcount\subsecno \global\subsecno=0
\def\subsec#1{\global\advance\subsecno by1\message{(\secsym\the\subsecno. #1)}
\ifnum\lastpenalty>9000\else\bigbreak\fi
\noindent{\it\secsym\the\subsecno. #1}\writetoca{\string\quad
{\secsym\the\subsecno.} {#1}}\par\nobreak\medskip\nobreak}
\def\appendix#1#2{\global\meqno=1\global\subsecno=0\xdef\secsym{\hbox{#1.}}
\bigbreak\bigskip\noindent{\bf Appendix #1. #2}\message{(#1. #2)}
\writetoca{Appendix {#1.} {#2}}\par\nobreak\medskip\nobreak}
%
%
\def\eqnn#1{\xdef #1{(\secsym\the\meqno)}\writedef{#1\leftbracket#1}%
\global\advance\meqno by1\wrlabeL#1}
\def\eqna#1{\xdef #1##1{\hbox{$(\secsym\the\meqno##1)$}}
\writedef{#1\numbersign1\leftbracket#1{\numbersign1}}%
\global\advance\meqno by1\wrlabeL{#1$\{\}$}}
\def\eqn#1#2{\xdef #1{(\secsym\the\meqno)}\writedef{#1\leftbracket#1}%
\global\advance\meqno by1$$#2\eqno#1\eqlabeL#1$$}
%
\newskip\footskip\footskip14pt plus 1pt minus 1pt 
\def\footnotefont{\ninepoint}\def\f@t#1{\footnotefont #1\@foot}
\def\f@@t{\baselineskip\footskip\bgroup\footnotefont\aftergroup\@foot\let\next}
\setbox\strutbox=\hbox{\vrule height9.5pt depth4.5pt width0pt}
\global\newcount\ftno \global\ftno=0
\def\foot{\global\advance\ftno by1\footnote{$^{\the\ftno}$}}
%
\newwrite\ftfile
\def\footend{\def\foot{\global\advance\ftno by1\chardef\wfile=\ftfile
$^{\the\ftno}$\ifnum\ftno=1\immediate\openout\ftfile=foots.tmp\fi%
\immediate\write\ftfile{\noexpand\smallskip%
\noexpand\item{f\the\ftno:\ }\pctsign}\findarg}%
\def\footatend{\vfill\eject\immediate\closeout\ftfile{\parindent=20pt
\centerline{\bf Footnotes}\nobreak\bigskip\input foots.tmp }}}
\def\footatend{}
%
%
\global\newcount\refno \global\refno=1
\newwrite\rfile
\def\ref{[\the\refno]\nref}
\def\nref#1{\xdef#1{[\the\refno]}\writedef{#1\leftbracket#1}%
\ifnum\refno=1\immediate\openout\rfile=refs.tmp\fi
\global\advance\refno by1\chardef\wfile=\rfile\immediate
\write\rfile{\noexpand\item{#1\ }\reflabeL{#1\hskip.31in}\pctsign}\findarg}
\def\findarg#1#{\begingroup\obeylines\newlinechar=`\^^M\pass@rg}
{\obeylines\gdef\pass@rg#1{\writ@line\relax #1^^M\hbox{}^^M}%
\gdef\writ@line#1^^M{\expandafter\toks0\expandafter{\striprel@x #1}%
\edef\next{\the\toks0}\ifx\next\em@rk\let\next=\endgroup\else\ifx\next\empty%
\else\immediate\write\wfile{\the\toks0}\fi\let\next=\writ@line\fi\next\relax}}
\def\striprel@x#1{} \def\em@rk{\hbox{}}
\def\lref{\begingroup\obeylines\lr@f}
\def\lr@f#1#2{\gdef#1{\ref#1{#2}}\endgroup\unskip}

\def\addref#1{\immediate\write\rfile{\noexpand\item{}#1}} 
\def\startrefs#1{\immediate\openout\rfile=refs.tmp\refno=#1}
\def\xref{\expandafter\xr@f}\def\xr@f[#1]{#1}
\def\refs#1{\count255=1[\r@fs #1{\hbox{}}]}
\def\r@fs#1{\ifx\und@fined#1\message{reflabel \string#1 is undefined.}%
\nref#1{need to supply reference \string#1.}\fi%
\vphantom{\hphantom{#1}}\edef\next{#1}\ifx\next\em@rk\def\next{}%
\else\ifx\next#1\ifodd\count255\relax\xref#1\count255=0\fi%
\else#1\count255=1\fi\let\next=\r@fs\fi\next}
%

%
\newwrite\ffile\global\newcount\figno \global\figno=1
\def\fig{fig.~\the\figno\nfig}
\def\nfig#1{\xdef#1{fig.~\the\figno}%
\writedef{#1\leftbracket fig.\noexpand~\the\figno}%
\ifnum\figno=1\immediate\openout\ffile=figs.tmp\fi\chardef\wfile=\ffile%
\immediate\write\ffile{\noexpand\medskip\noexpand\item{Fig.\ \the\figno. }
\reflabeL{#1\hskip.55in}\pctsign}\global\advance\figno by1\findarg}
\def\xfig{\expandafter\xf@g}\def\xf@g fig.\penalty\@M\ {}
\def\figs#1{figs.~\f@gs #1{\hbox{}}}
\def\f@gs#1{\edef\next{#1}\ifx\next\em@rk\def\next{}\else
\ifx\next#1\xfig #1\else#1\fi\let\next=\f@gs\fi\next}
\newwrite\lfile
{\escapechar-1\xdef\pctsign{\string\%}\xdef\leftbracket{\string\{}
\xdef\rightbracket{\string\}}\xdef\numbersign{\string\#}}

\def\writestop{\def\writestoppt{\immediate\write\lfile{\string\pageno%
\the\pageno\string\startrefs\leftbracket\the\refno\rightbracket%
\string\def\string\secsym\leftbracket\secsym\rightbracket%
\string\secno\the\secno\string\meqno\the\meqno}\immediate\closeout\lfile}}
\def\writestoppt{}\def\writedef#1{}
\def\seclab#1{\xdef #1{\the\secno}\writedef{#1\leftbracket#1}\wrlabeL{#1=#1}}
\def\subseclab#1{\xdef #1{\secsym\the\subsecno}%
\writedef{#1\leftbracket#1}\wrlabeL{#1=#1}}
\newwrite\tfile \def\writetoca#1{}
\def\leaderfill{\leaders\hbox to 1em{\hss.\hss}\hfill}
\def\writetoc{\immediate\openout\tfile=toc.tmp
   \def\writetoca##1{{\edef\next{\write\tfile{\noindent ##1
   \string\leaderfill {\noexpand\number\pageno} \par}}\next}}}

%
\catcode`\@=12 
%
\edef\tfontsize{\ifx\answ\bigans scaled\magstep3\else scaled\magstep4\fi}
\font\titlerm=cmr10 \tfontsize \font\titlerms=cmr7 \tfontsize
\font\titlermss=cmr5 \tfontsize \font\titlei=cmmi10 \tfontsize
\font\titleis=cmmi7 \tfontsize \font\titleiss=cmmi5 \tfontsize
\font\titlesy=cmsy10 \tfontsize \font\titlesys=cmsy7 \tfontsize
\font\titlesyss=cmsy5 \tfontsize \font\titleit=cmti10 \tfontsize
\skewchar\titlei='177 \skewchar\titleis='177 \skewchar\titleiss='177
\skewchar\titlesy='60 \skewchar\titlesys='60 \skewchar\titlesyss='60
\def\titlefont{\def\rm{\fam0\titlerm}
\textfont0=\titlerm \scriptfont0=\titlerms \scriptscriptfont0=\titlermss
\textfont1=\titlei \scriptfont1=\titleis \scriptscriptfont1=\titleiss
\textfont2=\titlesy \scriptfont2=\titlesys \scriptscriptfont2=\titlesyss
\textfont\itfam=\titleit \def\it{\fam\itfam\titleit}\rm}
 \ifx\answ\bigans\else scaled\magstep1\fi
\ifx\answ\bigans\else

 \font\absi=cmmi10 scaled\magstep1
\font\absis=cmmi7 scaled\magstep1 \font\absiss=cmmi5 scaled\magstep1
\font\abssy=cmsy10 scaled\magstep1 \font\abssys=cmsy7 scaled\magstep1
\font\abssyss=cmsy5 scaled\magstep1 
\skewchar\absi='177 \skewchar\absis='177 \skewchar\absiss='177
\skewchar\abssy='60 \skewchar\abssys='60 \skewchar\abssyss='60
\fi
\font\ninerm=cmr9 \font\sixrm=cmr6 \font\ninei=cmmi9 \font\sixi=cmmi6
\font\ninesy=cmsy9 \font\sixsy=cmsy6 \font\ninebf=cmbx9
\font\nineit=cmti9 \font\ninesl=cmsl9 \skewchar\ninei='177
\skewchar\sixi='177 \skewchar\ninesy='60 \skewchar\sixsy='60
\def\ninepoint{\def\rm{\fam0\ninerm}
\textfont0=\ninerm \scriptfont0=\sixrm \scriptscriptfont0=\fiverm
\textfont1=\ninei \scriptfont1=\sixi \scriptscriptfont1=\fivei
\textfont2=\ninesy \scriptfont2=\sixsy \scriptscriptfont2=\fivesy
\textfont\itfam=\ninei \def\it{\fam\itfam\nineit}\def\sl{\fam\slfam\ninesl}%
\textfont\bffam=\ninebf \def\bf{\fam\bffam\ninebf}\rm}
%
%

\hyphenation{anom-aly anom-alies coun-ter-term coun-ter-terms}
\def\inv{^{\raise.15ex\hbox{${\scriptscriptstyle -}$}\kern-.05em 1}}

\def\Dsl{\,\raise.15ex\hbox{/}\mkern-13.5mu D} 
\def\dsl{\raise.15ex\hbox{/}\kern-.57em\partial}
\def\del{\partial}

\def\lspace{\ifx\answ\bigans{}\else\qquad\fi}
\def\lbspace{\ifx\answ\bigans{}\else\hskip-.2in\fi} 
\def\boxeqn#1{\vcenter{\vbox{\hrule\hbox{\vrule\kern3pt\vbox{\kern3pt
	\hbox{${\displaystyle #1}$}\kern3pt}\kern3pt\vrule}\hrule}}}
\def\mbox#1#2{\vcenter{\hrule \hbox{\vrule height#2in
		\kern#1in \vrule} \hrule}}  
%

\def\darr#1{\raise1.5ex\hbox{$\leftrightarrow$}\mkern-16.5mu #1}

\def\roughly#1{\raise.3ex\hbox{$#1$\kern-.75em\lower1ex\hbox{$\sim$}}}

\def\complex{{ {\rm C} \kern -.5em
   {\raise .12ex \hbox{\vrule height 1.2ex width 0.02em depth 0ex}}
   \kern 0.5em }}
\def\real{{ {\rm R} \kern -0.45em \vrule height 1.44ex width 0.02em depth 0ex
   \kern 0.45em }}
\def\tto{{ \;\rightarrow\; }}
\def\sn{{\; \hbox{sn}\, }}
\def\cn{{\; \hbox{cn}\, }}
\def\dn{{\; \hbox{dn}\, }}
\def\nc{{\; \hbox{nc}\, }}
\def\sc{{\; \hbox{sc}\, }}

{\nopagenumbers
\rightline{UGVA-DPT-1994/05-853}
\rightline{MIT-CTP-2328}
\rightline{June 1994}
\bigskip\bigskip
\centerline{\titlefont Lattice Poincar\'e as a quantum deformed algebra}
\bigskip
\centerline{Cesar Gomez
\foot{On leave of absence from CSIC, c/ Serrano 119, 28006 Madrid,
Spain.}$^{,2}$,
Henri Ruegg
\foot{Partially supported by the Swiss National Science Foundation and
OFES Contract No 93.0083,
Human Capital and Mobility Contract No ERBCHRXCT920069.}
and Philippe Zaugg
\foot{Present address:  Center for Theoretical Physics, MIT,
77 Massachusetts Avenue, Cambridge, MA 02139, USA.
Partially supported by the Swiss National Science Foundation
and by funds provided by the U.S. Department of Energy (D.O.E.) under
cooperative agreement \#DE-FC02-94ER40818.} }
\bigskip
\centerline{\sl D\'epartement de Physique th\'eorique}
\centerline{\sl Universit\'e de Gen\`eve}
\centerline{\sl 32, bd d'Yvoy}
\centerline{\sl CH--1211 Gen\`eve 4}
\centerline{\sl Switzerland}
\vskip 1in
\centerline{\bf Abstract}
\medskip
We propose a definition of a Poincar\'e algebra for a two dimensional
space--time with one discretized dimension. This algebra has the structure
of a Hopf algebra. We use the link between Onsager's uniformization
of the Ising model and the dispersion relation of a free particle in this
space--time, together with the rapidity representation of the quantum
deformation of the Poincar\'e enveloping algebra.
\vfill
\pageno=0
\eject}
\ftno=0
\newsec{Lattice Poincar\'e}

Lattice generalizations of the Poincar\'e or inhomogeneous Lorentz invariance
have been proposed in the context of two dimensional integrable models
\ref\tetel{M.~G.~Tetel'man,
Sov. Phys. JETP {\bf 55} (2) (1982), 306.}
\ref\thacker{H.~B.~Thacker,
Physica {\bf 18D} (1986) 348.}.
These works were inspired by Onsager's solution of the Ising model
\ref\ons{L.~Onsager,
Phys. Rev. {\bf D16} (1944) 117.}
and Baxter's definition of the corner transfer matrix
\ref\bax{R.~J.~Baxter, J. Stat. Phys. {\bf 15} (1976), 485.}.

Onsager's solution to the Ising model provides a natural way to associate a
continuous rapidity to a free massive fermion living in a discrete two
dimensional space--time. In fact, mapping the lattice spacing $a_x, a_t$
\thacker\ into the Ising couplings $H$ and $H'$, $H^*$ being the dual of $H$
\ons,
\eqn\couplings{
\eqalign{
  \sinh 2H' \sinh 2H^* & = \left({a_x \over a_t}\right)^2 \cr
  2 \sinh(H'-H^*) & = \mu a_x } }
and defining
\eqn\rap{
\eqalign{
\gamma = p a_x \cr
\omega = E a_t } }
we get, from Onsager's hypergeometric relation
\eqn\hyper{
\cosh \gamma = \cosh 2H' \cosh 2H^* - \sinh 2H' \sinh 2H^* \cos \omega ,}
the lattice dispersion relation:
\eqn\disp{
a_t^2 ( \cosh p a_x -1) + a_x^2 (\cos E a_t -1) = {\mu^2 \over 2} a^2_t a^2_x
.}

Now, using Onsager's uniformization of equation \hyper, we get the desired
rapidity for the free particles moving in a discrete two dimensional
space--time. For fixed Ising couplings $H, H'$ the rapidity $\alpha$
is defined by the following uniformization relations in terms of Jacobi
elliptic functions:
\eqn\sinus{
\eqalign{
& \sinh 2 H' = -i \sn (i a | k^2) \cr
& \sinh 2 H^* = -i k \sn (i a | k^2) \cr
& \sinh \gamma = -i{1-k^2 \over M} \sn (i a | k^2) \cr
&\sin \omega = {1-k^2 \over M} \sn (\alpha | k^2) } }
with the short hand
\eqn\mm{
M = \dn(i a | k^2) \dn(\alpha | k^2) + k \cn(i a | k^2) \cn(\alpha| k^2) . }
The elliptic modulus $k$ is defined by the Ising
integrability relation
\eqn\integ{
k = {\sinh 2 H^* \over \sinh 2 H'} }
and for $H' > H^*$, $k<1$.
As usual, integrability means the commutativity of the transfer
matrix for two different values of the rapidity $\alpha$.

We now introduce the substitution \couplings\ into \sinus\ for an
uniformization of the dispersion relation in term of the lattice variables
$a_x, a_t, p, E$ and obtain:
\eqn\truc{
\eqalign{
& -i \sn( i a| k^2) = {1 \over \sqrt{ k}} {a_x \over a_t} \cr
& M = {\sqrt{k} \over a_t} \left( \sqrt{{k a_x^2+a_t^2 \over k}}
\dn(\alpha| k^2) + \sqrt{a_x^2 + k a_t^2} \cn(\alpha| k^2) \right) \cr
& k_\pm = 1 + A \pm \sqrt{2 A + A^2} \cr
& A = { \mu^2(a_t^2 + a_x^2) \over 2} + {\mu^4 a_t^2 a_x^2 \over 8} } }
In the continumm limit where both $a_x, a_t \tto 0$, and therefore $k \tto 1$,
equations
\rap, together with \sinus, become
\eqn\relativ{
p = \mu \cosh \alpha , \qquad \qquad E = \mu \sinh \alpha }
which shows that $\alpha$ is the rapidity, with the mass shell condition
$p^2 -E^2 = \mu^2$ (although usually $E$ and $p$
are interchanged). This suggests to define a boost generator $N$ by
\eqn\boost{
N = {\del \over \del \alpha} .}

The relations \relativ\ can be obtained  as solutions to the differential
equations
\eqn\eqdiff{
\eqalign{
 {\del P(\alpha) \over \del \alpha} & = E(\alpha) \cr
 {\del^2 P(\alpha) \over \del \alpha^2} & = P(\alpha) } }
which are equivalent  to the standard two dimensional Poincar\'e algebra in
the continuum:
\eqn\contP{
\eqalign{
  [ N,P] & = E \cr
  [ N,E] & = P \cr
  [E,P] & = 0 } }
once we use the rapidity representation \boost\ for the boost generator.

In order to define the lattice generalization of the two dimensional
Poincar\'e algebra, we should modify the algebra \contP\ in such a way that
the lattice rapidity representation obtained using Onsager's uniformization
\sinus\ appears now as a solution to the differential equations defined by
the modified algebra.

In this letter we prove that at least for a discrete space and a continuous
time ($a_t \tto 0$), the lattice generalization of \contP\ is a
Hopf algebra possessing an asymmetric comultiplication for $N$ and $E$.

\newsec{From lattice Poincar\'e to quantum Poincar\'e}

A quantum deformation of the Poincar\'e algebra
has been recently introduced in
\ref\italy{E.~Celeghini, R.~Giachetti, E.~Sorace and M.~Tarlini, J. Math. Phys.
{\bf 31} (1990) 2548; {\bf 32} (1991) 1155, 1159.\hfill\break
F.~Bonechi, E.~Celeghini, R.~Giachetti, E.~Sorace and M.~Tarlini, Phys.
Rev. Lett.{\bf 68} (1992) 3718.}
\ref\lnr{J.~Lukierski, A.~Nowicki, H.~Ruegg and V.~Tolstoy, Phys. Lett. {\bf
B264} (1991) 331.\hfill\break
J.~Lukierski, A.~Nowicki, H.~Ruegg, Phys. Lett. {\bf B293} (1992) 344.}.
For two dimensions the $\kappa$--Poincar\'e algebra $U_\kappa(P_2)$ is
defined by the commutation relations:
\eqn\kptwo{
\eqalign{
[N,P] &= E \cr
[N,E] &= \kappa \sinh {P \over \kappa} \cr
[E,P] &= 0 } }
with $\kappa \in \real$. Introducing a rapidity $\eta$ we get from \kptwo:
\eqn\kpdiff{
\eqalign{
{\del P(\eta) \over \del \eta} & = E(\eta) \cr
{\del^2 P(\eta) \over \del \eta^2} & = \kappa \sinh {P(\eta) \over \kappa} }}
It is now easy to prove that the semi-continuous limit $a_t \tto 0$ of
equations \sinus\ defines a solution to the equations \kpdiff.
In fact solving \kpdiff\ we get
\ref\rrr{J.~Lukierski, H.~Ruegg and W.~R\" uhl, Phys. Lett. {\bf B313} (1993)
357.}:
\eqn\kpsol{
\eqalign{
2 \kappa \sinh {P(\eta) \over 2 \kappa} =  \mu \nc(K_L^{-1} \eta | K^2_L) \cr
E(\eta) = \mu \sc( K_L^{-1} \eta | K^2_L ) } }
with the elliptic modulus
\eqn\defk{
K^2_L = {1 \over 1 + \mu^2/4 \kappa^2} .}
On the other hand, taking the $a_t \tto 0$ limit of \sinus, we get:
\eqn\sinuslim{
\eqalign{
\sinh \gamma &= {1-k^2 \over k} {1 \over \cn \alpha + \dn \alpha} \cr
\sin \omega & \simeq E a_t = {a_t \over a_x} {1-k^2 \over k} {\sn \alpha
 \over \cn \alpha + \dn \alpha} } }
whereby
\eqn\energy{
E= {1 \over a_x} {1-k^2 \over k} {\sn \alpha
 \over \cn \alpha + \dn \alpha} .}
By performing a Landen transformation
\ref\table{M.~Abramowitz and I.~A.~Stegun, {\sl Handbook of mathematical
functions}, Appl. Math. Series 55 (1964) National Bureau of Standards, USA.}
we can identify \kpsol\ and \sinuslim, provided:
\eqn\matching{
\sqrt{k} \alpha = \eta \qquad\hbox{and}\qquad a_x = {1 \over \kappa} .}

Therefore we conclude that in the semi-continuous limit the two dimensional
lattice Poincar\'e algebra is defined by the quantum deformation \kptwo, with
the deformation parameter $\kappa$ being determined by the lattice spacing.
As an extra piece of evidence, notice that the lattice dispersion relation
\disp\ becomes in the $a_t \tto 0$ limit the Casimir
\eqn\casimir{
C = \left( 2 \kappa \sinh {p \over 2 \kappa} \right)^2 - E^2 = \mu^2 }
of the quantum Poincar\'e algebra \kptwo.

In the continuum limit $a_x \tto 0$, (i.e. $\kappa \tto \infty$), the elliptic
parameter $k$ becomes 1 \truc, which implies $H'=H^*$ in \integ, i.e. $T=T_c$.
Hence $\kappa$ measures the departure from criticality.

The algebra \kptwo\ can be promoted to a Hopf algebra by the following
comultiplication rules:
\eqn\comult{
\eqalign{
\Delta P &= P \otimes 1 + 1 \otimes P \cr
\Delta E &= E \otimes e^{P/2\kappa} + e^{-P/2\kappa} \otimes E \cr
\Delta N &= N \otimes e^{P/2\kappa} + e^{-P/2\kappa} \otimes N } }
and antipode
\eqn\antipode{
S(P) = -P, \qquad S(E) = -E, \qquad S(N) = -N +E/2\kappa,}
as well as a trivial counit $\epsilon(X)=0$ for $X=P,E,N$.
An interesting feature is
that for finite lattice spacing, the comultiplication for $E$ and $N$
are asymmetric.

According to \comult, the total energy and momemtum for a
system of two particles are given by:
\eqn\sumEP{
\eqalign{
P^t &= P_1 + P_2 \cr
E^t &= E_1  e^{P_2/2\kappa} +e^{-P_1/2\kappa} E_2 }}
satisfying the Casimir relation
\eqn\casimirtwo{
\hbox{cst} = \left( 2 \kappa \sinh {P^t \over 2 \kappa} \right)^2 -
(E^t)^2  .}

It should be noted that the comultiplication \comult\ is not unique.
The algebra \kptwo\ is compatible with a symmetric comultiplication which is
additive for $E$, non--additive for $P$ and complicated for $N$
\ref\pawel{P.~Maslanka, J. Math. Phys. {\bf 34} (1993) 6025.}.
\newsec{Final comments}

The main goal of this short note was to provide a definition of the lattice
Poincar\'e algebra in terms of well defined differential operators. The main
tool that we used, inspired by integrable models, was the introduction of
a continuous but elliptic rapidity, translating in that way the lattice
information into the elliptic parameter. More precisely, the steps followed
in defining this algebra were:

1) To uniformize the lattice dispersion relation, for a free massive particle,
by means of Onsager's uniformization of the Ising model. The reason we used the
Ising model is simply because of its equivalence to a free fermionic system.

2) To represent the boost generator of the Poincar\'e algebra as the derivative
with respect to the rapidity variable, which is identified with the
elliptic uniformization parameter.

3) To integrate, using the elliptic rapidity, the lattice dispersion relation
\disp in the semi-continuum limit.  In this way we arrive to equations \kpdiff,
which define the
quantum Poincar\'e algebra introduced in \italy\ \lnr. In this algebra,
the only memory of its lattice origin is the relation \matching\ between the
quantum deformation parameter and the lattice spacing.

The unexpected output of this exercise, which was done on the basis of one
body information, is the kinematical implication on the many body dynamics
which is encoded in the Hopf algebra structure (comultiplication rules)
of the quantum (lattice) Poincar\'e algebra. The non triviality of the
comultiplication rules directly derives from the non trivial topology of
the rapidity space, namely we have lost two conformal Killing vectors. The
situation is somewhat similar to the standard analogy between quantum
$SU(2)$ and the asymmetric top.

The physical meaning of the comultiplication
rules becomes tricky when we deal with a free system. In general, in order to
define the free Hamiltonian for $n$ particles, we just add $n$ free one
body Hamiltonians, which is equivalent to assume trivial comultiplication
rules for kinematical observables. Therefore we observe that the quantum
deformation of any kinematical symmetry requires the existence of some
non local interactions, to account for the non trivial comultiplication.

There are two physical questions that deserve some attention at this point.
The first one concerns the interplay between integrability and lattice
kinematics, interpreted in the way proposed in this note. For an integrable
model in the elliptic regime we can always try to define an  associated
Poincar\'e algebra where the rapidity is the uniformization parameter and
the boost generator corresponds to the corner transfer matrix.
{}From this kinematical point of view, integrability becomes equivalent to
the relativity principle, i.e. physics is the same for two observers
related by a boost, which corresponds to the commutativity of the transfer
matrices for two different values of the uniformization parameter. The
Poincar\'e algebra so obtained would be very likely a quantum deformed
algebra with the quantum deformation parameter being determined by the
integrability constants (the parameters determining the integrability
manifold in the space of couplings) of the model. It would be very
interesting, if the previous picture is correct, to find the physical
interpretation in this context of the comultiplication rules.
It should be noted that in previous references \tetel\ \thacker\ a lattice
Poincar\'e was defined, for integrable models, using the infinite set
of conserved charges. In the lattice Poincar\'e considered here,
we play with only three generators but we are forced to work, as is
customary for quantum deformations, with the enveloping algebra.

The second question concerns the interpretation of the quantum Poincar\'e
algebra as a lattice regularization. From the two dimensional
semi-continuous case studied here, we learn that the lattice spacing
transmutes into a quantum deformation parameter modifying automatically
the comultiplication rules. A possible physical picture for understanding
this phenomena is to interpret the non trivial comultiplication rules as
a way to save formally the conservation laws lost when postulating a
fundamental length. An interesting exercise in this direction would be
to derive, from the deformed Poincar\'e algebra and for two dimensional
models, a germ (i.e. the M\" obius part) of a deformed conformal algebra useful
away from criticality.

To conclude we just want to point out that the perhaps most pragmatic and
useful approach to quantum deformed kinematics is to interpret it as a subtle
way of lattice regularization.
\newsec{Acknowledgments}

We thank M. Ruiz--Altaba for discussions,
V.~Rittenberg for critical comments and M.~Henkel for discussions.
\footatend\vfill\supereject\immediate\closeout\rfile\writestoppt
\baselineskip=14pt\centerline{{\bf References}}\bigskip{\frenchspacing%
\parindent=20pt\escapechar=` \input refs.tmp\vfill\eject}\nonfrenchspacing
\bye